\documentclass[%
reprint,
showpacs,preprintnumbers,
 amsmath,amssymb,
 pre,
]{revtex4-1}

\usepackage{graphicx}% Include figure files
\usepackage{dcolumn}% Align table columns on decimal point
\usepackage{bm}% bold math
\newcommand{\eref}[1]{Eq.~(\ref{eq:#1})}
\newcommand{\fref}[1]{Fig.~\ref{fig:#1}}
\newcommand{\sref}[1]{Sec.~\ref{sec:#1}}

\newcommand{\lr}[1]{ \left( #1 \right) }
\newcommand{\LR}[1]{ \left[ #1 \right] }
\newcommand{\llrr}[1]{ \left\{ #1 \right\} }
\newcommand{\av}[1]{ \left< #1 \right> }

\renewcommand{\vec}[1]{ \mathbf{#1} }

\newcommand{\position}{\vec{r}}

\newcommand{\force}{\vec{F}}
\newcommand{\shearrate}{\dot{\gamma}}
\newcommand{\sedimentationStrength}{ \epsilon_{sed} }

\usepackage{color}

\begin{document}

\preprint{APS/123-QED}

\title{Depinning and heterogeneous dynamics of colloidal crystal layers under shear flow}% Force line breaks with \\

\author{Sascha Gerloff}
\email{s.gerloff@tu-berlin.de}
\author{Sabine H. L. Klapp}
\email{klapp@physik.tu-berlin.de}
\affiliation{%
 Institut f\"ur Theoretische Physik, Hardenbergstr. 36,\\ Technische Universit\"at Berlin, D-10623 Berlin, Germany
}%

\date{\today}% It is always \today, today,

\begin{abstract}
Using Brownian dynamics (BD) simulations and an analytical approach we investigate the shear-induced, nonequilibrium dynamics of dense colloidal suspensions confined to a narrow slit-pore.
Focusing on situations where the colloids arrange in well-defined layers with solidlike in-plane structure, the confined films display complex, nonlinear behavior such as collective depinning and local transport via density excitations.
These phenomena are reminiscent of colloidal monolayers driven over a periodic substrate potential.
In order to deepen this connection, we present an effective model which maps the dynamics of the shear-driven colloidal layers to the motion of a single particle driven over an effective substrate potential.
This model allows to estimate the critical shear rate of the depinning transition based on the equilibrium configuration, revealing the impact of important parameters such as the slit-pore width and the interaction strength.
We then turn to heterogeneous systems where a layer of small colloids is sheared with respect to bottom layers of large particles.
For these incommensurate systems we find that the particle transport is dominated by density excitations resembling the so-called "kink" solutions of the Frenkel-Kontorova (FK) model.
In contrast to the FK model, however, the corresponding "antikinks" do not move.
\end{abstract}
%
%\pacs{82.70.Dd, 05.70.Ln}% PACS, the Physics and Astronomy
\maketitle
%
%\tableofcontents
%
\section{\label{sec:introduction}Introduction}
Understanding the nonlinear response of dense colloidal systems to shear or other mechanical driving forces on a microscopic (i.e., particle-resolved) level has become a focus of growing interest.
Recent examples include density excitations (determining frictional properties) in driven colloidal monolayers \cite{Bohlein2012,Hasnain2013,Vanossi2012,Vanossi2013}, the stick-slip motion involved in the transmission of torque in driven colloidal clutches \cite{Williams2016}, as well as heterogeneities \cite{Benzi2014, Chaudhuri2013, Hentschel2016,Swayamjyoti2016},
and diverging stress- and strain correlations \cite{Benzi2014, Nicolas2014} in sheared colloidal glasses.
Related complex microscopic behavior occurs in sheared granular matter \cite{Denisov2016} and sheared suspensions of non-Brownian particles \cite{Fornari2016}.
Developing a microscopic understanding of such shear-induced behavior is interesting not only in the general context of nonequilibrium behavior of soft-matter systems, but also is crucial for applications in nanotribology, the design of novel materials and of efficient nanomachines.

In the present paper we are concerned with the shear-induced microscopic response of thin films of spherical colloidal particles between two planar walls (slit-pore geometry).
By using Brownian Dynamics (BD) computer simulations and an analytical approach, we aim at understanding transport mechanisms under shear for both, mono- and bidisperse systems.

The structural behavior of colloidal suspensions in presence of spatial confinement is nontrivial already in equilibrium; in particular, it is well established that the particles spontaneously form layers (see, e.g., \cite{Klapp2008}) which, moreover, become crystal-like ("capillary freezing") in lateral directions at sufficiently high densities \cite{Grandner2010}.
Exposing such highly correlated systems to shear flow (along a direction within the plane of the walls) leads to a breakdown of crystalline in-plane ordering after overcoming a "critical" shear rate, and a subsequent recrystallization at higher shear rates, as both computer simulations \cite{Messina2006, Vezirov2013} and experiments \cite{Reinmueller2013} reveal.
In two earlier publications \cite{Vezirov2013,Vezirov2015} we have analyzed this behavior in detail, for the exemplary case of a colloidal bilayer (of monodisperse particles) under constant shear rate \cite{Vezirov2013} or constant stress \cite{Vezirov2015} (both of these external control strategies can be experimentally realized).
One main conclusion was that the breakdown of crystalline order is related to "depinning" transitions in terms of the layer velocity from a locked into a running (sliding) state \cite{Vezirov2013}.
In this sense, the dynamical behavior of confined colloidal layers under shear bears strong similarities to the well-studied case of one-dimensional (1D) particle chains or two-dimensional (2D) particle monolayers driven over a periodic substrate \cite{Hasnain2014, Reimann2002, Risken1996}.

Inspired by this similarity, we here propose an analytical model which allows to predict the shear-induced depinning on the basis of the structure in thermal equilibrium.
The model is essentially a variant of the well-known Frenkel-Kontorova (FK) model \cite{Braun1998,Kontorova1939}, which has been extensively used to model friction between solid (atomic or colloidal) surfaces and has also proven to be crucial for understanding driven monolayers \cite{Bohlein2012,Hasnain2013}.
It should be stressed, however, that despite all similarities, there is one crucial difference between our system and the case of driven monolayers: in the latter case, the periodic substrate represents a fixed \emph{external field}, whereas in our case, the "substrate" rather corresponds to a neighboring layer which can respond to the shear flow itself by in- and out-of-plane deformations.
Indeed, one main goal of the present study is to elucidate the implications of this difference.

A further major goal is to explore the impact of incommensurability, that is, a mismatch of structural length scales, in our sheared system. To this end we consider an asymmetric system where a layer of small colloids is sheared with respect to (crystalline) layers of larger particles.
As expected from the FK model as well as from previous, experimental \cite{Bohlein2012} and theoretical \cite{Hasnain2013, McDermott2013, Siems2015} studies of driven monolayers, we observe moving defect structures with locally enhanced density ("kinks") or locally reduced density ("antikinks").
These kinks and antikinks correspond to soliton solutions of the continuum version of the FK model (i.e., the sine-Gordon equation).
Contrary to the theoretically predicted scenario, however, in our system only the kinks participate in the particle transport, whereas the antikinks remain essentially "locked" within the moving layer.

The rest of the papers is organized as follows. In \sref{models and simulation details} we describe our (mono- or bidisperse) model systems and the details of our BD simulations.
In \sref{average motion} we give a first overview of the behavior of the different films by considering simulation results for
the average motion of the layers.
We then proceed by presenting our analytical model which targets mainly the bilayer system (in \sref{bilayer}). However, we also discuss its application to a monodisperse trilayer system (in \sref{trilayer}).
Section~\ref{sec:density excitations} is devoted to the bidisperse system, for which we discuss in detail the local transport via density excitations.
We close with a summary and conclusion in \sref{conclusion}.
\section{\label{sec:models and simulation details}Models and simulation details}
\subsection{\label{sec:model systems}Model systems}
We consider a colloidal suspension consisting of macroions of diameter $d_i$, salt ions, counterions, and solvent molecules.
Focusing on the macroions, the influence of the solvent is considered implicitly by employing the Derjaguin-Landau-Verwey-Overbeek (DLVO) approximation.
In this framework, the electrostatic interaction of the macroions is screened by the salt- and counterions leading (on a mean-field level) to a \emph{Yukawa}-like potential
\begin{equation}\label{eq:dlvo-potential}
 U_{ \text{DLVO} }(r_{ij}) = V_{ij} \frac{\exp(-\kappa \, r_{ij})}{r_{ij}},
\end{equation}
with the pair interaction strength $V_{ij}$, the inverse Debye screening length $\kappa$, and the particle distance $r_{ij}$.
The interaction parameters are set in accordance to real suspensions of charged silica particles with a diameter of about $d \approx 26\,nm$ \cite{Klapp2007}, yielding $\kappa d \approx 3.2$.
In order to account for the steric repulsion between the macroions we supplement the DLVO potential by a soft-sphere (SS) potential, which is given by the repulsive part of the Lennard-Jones potential
\begin{equation}\label{eq:soft-sphere potential}
 U_{ \text{SS} }(r_{ij}) = 4\epsilon_{ \text{SS} }\lr{\frac{d_{ij} }{r_{ij} } }^{12},
\end{equation}
with the interaction strength $\epsilon_{ \text{SS} }$ and the mean particle diameter $d_{ij} = \lr{d_i+d_j}/2$.
Therefore, the total particle interaction between two macroions reads
\begin{equation}\label{eq:total particle interaction}
 U_{inter}(r_{ij}) = U_{ \text{DLVO} }(r_{ij}) + U_{ \text{SS} }(r_{ij}).
\end{equation}
Following previous studies, the total particle interaction potential is truncated at a cutoff radius $r_{c} \approx 3d$ and shifted accordingly \cite{Vezirov2013,Vezirov2015}.

To mimic the slit-pore geometry, the colloids are confined by two plane-parallel soft walls extended infinitely in $x$- and $y$-direction and separated in $z$-direction by a distance $L_z$ (see \fref{sketch}). The interaction between the colloids and the walls is described by
\begin{equation}\label{eq:soft-wall potential}
 U_{wall}(z_i) = \frac{4\pi\epsilon_w}{5} \LR{ \left( \frac{d_{i,w} }{ L_z/2 - z_i } \right)^9  + \left( \frac{d_{i,w} }{ L_z/2 + z_i } \right)^9 },
\end{equation}
with $z_i$ being the $z$-coordinate of particle $i$, the mean wall diameter $d_{i,w} = (d_i+d_w)/2$, the wall diameter $d_w = d$, and the wall-interaction strength $\epsilon_w$.
Equation~(\ref{eq:soft-wall potential}) is obtained by integrating over a half-space of continuously distributed uncharged soft wall-particles, where the interaction between the wall- and the fluid particles is set to the repulsive part of the Lennard-Jones potential [see \eref{soft-sphere potential} with diameter $d_w$].
It is widely adopted as a model for the fluid-wall interaction \cite{Jean-PierreHansen2013, Klapp2007}.\\

In this study, we focus on systems where $L_z$ is of the order of the particle diameter and the density is rather high.
In such situations the colloids arrange in well-defined layers with a solidlike in-plane structure (at least in equilibrium).
Further, we consider both, one-component systems and a special type of a binary mixture.
The latter involves particles with two different diameters, the idea being to create a structure with a mismatch of the underlying structural length scales of the corresponding pure systems.
Specifically, we aim to create a structure where small colloidal particles form a "top" layer on a crystal of larger particles.
In order to stabilize such an asymmetric situation (which would not arise with a symmetric external potential), we supplement the confinement potential by a linear "sedimentation" potential
\begin{equation}\label{eq:sedimentation potential}
 U_{sed}(z_i) = \sedimentationStrength d_i^3 z_i,
\end{equation}
with the sedimentation potential strength $\sedimentationStrength$.
This potential can be formally interpreted as the first-order term of a Taylor expansion of the gravitational potential in $z_i$ \cite{Cuetos2006, Ginot2015, Heras2016}.
The resulting force $F_{sed}\lr{\position_i} = -\nabla_{\position_i} U_{sed}\lr{z_i}$ depends for constant $\sedimentationStrength$ only on the diameter $d_i$ of the particle $i$ and therefore leads to the sedimentation of large colloids.
For appropriate values of $\sedimentationStrength$ we find stable configurations consisting of crystalline layers of large particles at the bottom and a layer of small particles on top.
\subsection{\label{sec:simulation details}Simulation details}
\begin{figure}
  \includegraphics[width=1.0\linewidth,natheight=1102,natwidth=2102]{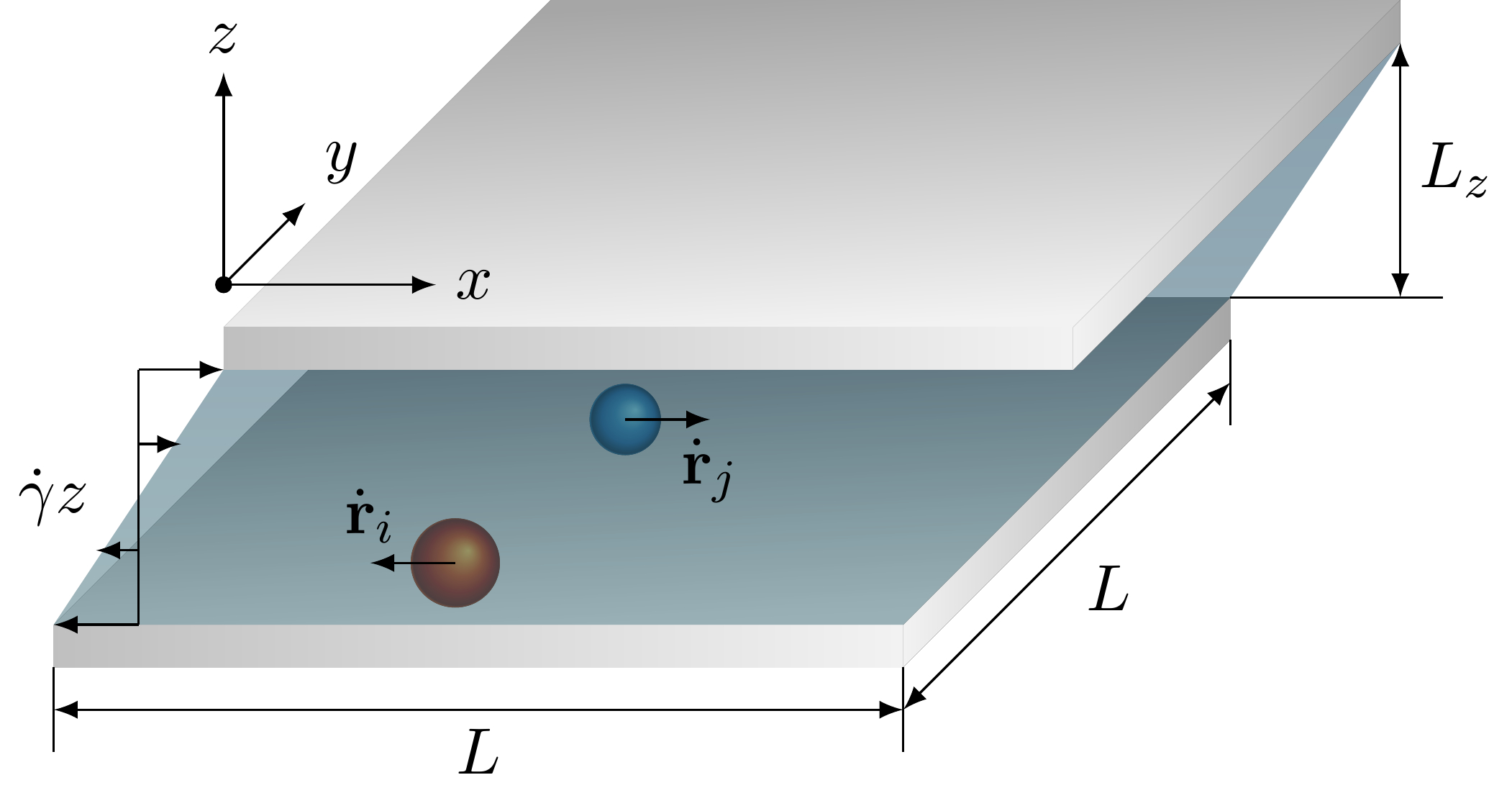}
	\caption{(Color online) Sketch of the model system, involving colloidal particles in narrow slit-pore confinement and linear shear flow in $x$-direction with gradient $\shearrate z$ in $z$-direction. Periodic boundary conditions are applied in $x$- and $y$-direction. The width of the slit-pore is set to $L_z$.}
	\label{fig:sketch}
\end{figure}
We perform standard (overdamped) BD simulations to examine the nonequilibrium properties and dynamics of our model systems.
The position $\position_i$ of particle $i$ is advanced according to the equation of motion \cite{Ermak1975}
\begin{equation}\label{eq:equation of motion}
 \position_i \lr{ t + \delta t } = \position_i \lr{ t } +\mu \force_i \lr{ \llrr{ \position } } \delta t + \delta \vec{W}_i + \shearrate z_i \delta t \vec{e}_x,
\end{equation}
where $\force_i$ is the total conservative force
(stemming from two-particle interactions [see \eref{total particle interaction}], particle-wall interactions [see \eref{soft-wall potential}], and the sedimentation potential [see \eref{sedimentation potential}])
acting on particle $i$, $\llrr{ \position } = \position_1 ,\ldots, \position_N $ is the set of particle positions, and $\delta t$ is the time step.
Within the framework of BD, the influence of the solvent is mimicked by a single-particle frictional- and random force.
The inverse friction constant defines the mobility $\mu= D_0 /k_BT$, where $D_0$ is the short-time diffusion coefficient, $k_B$ is the Boltzmann constant, and $T$ is the temperature.
The random force is modeled by random Gaussian displacements $\delta \vec{W}_i$ with zero mean and variance $2D_0 \delta t$ for each Cartesian component.
The timescale of the system was set to $\tau=d^2/D_0$, which defines the so-called Brownian time.
We impose a linear shear profile $ \shearrate z_i \vec{e}_x $ [see last term in \eref{equation of motion}] representing flow in $x$- and gradient in $z$-direction.
The strength of the flow is characterized by the uniform shear rate $\shearrate$.
This ansatz seems plausible for systems where the impact of the walls on the driving mechanism can be neglected, such as charged colloids confined between likewise charged, smooth walls \cite{Klapp2007, Reinmueller2013a}.
For this situation, the distance between the colloids and the wall is naturally rather large, suggesting that the motion of the colloids is not directly coupled to that of the particles comprising the wall.
Thus, one may assume that the shear flow away from the wall is approximately linear.
We note that, despite the application of a linear shear profile, the real, steady-state flow profile can be nonlinear \cite{Delhommelle2003}.
The present simulation approach has also been employed in other recent simulation studies of sheared colloids \cite{Besseling2012, Cerda2008, Lander2013};
the same holds for the fact that we neglect hydrodynamic interactions.
Furthermore, similar approaches have been employed in shear flow simulations of polymers at an interface \citep{Kekre2010,Radtke2014} and active particles in confinement \cite{Apaza2016}.

For the one-component bilayer- and trilayer system, the number density $\rho d^3 = 0.85$ and the slit-pore width $L_z = 2.2d, 3.2d$ are chosen following previous studies \cite{Vezirov2013,Vezirov2015}.
The particle interaction parameters are set according to experimental setups for particles with diameter $d \approx 26\,nm$ and valency $Z=35$ \cite{Zeng2011,Klapp2007}, yielding $\kappa d\approx3.2$.
For the two-component system, an additional small particle species is introduced with diameter $d_2 = 0.42d$ and valency $Z_2 = 0.17Z$, which are set according to experimental setups \cite{Zeng2011}, where we set $\kappa d \approx 3.3$ for all particles.
The number density $\rho d^3 = 1.226$ and the slit-pore width $L_z = 2.65d$ of the two-component system are chosen such that the volume density is comparable to the one-component system.
The sedimentation potential strength is set to $\sedimentationStrength=300k_BT/d^4$ for the two-component system and zero for all one-component systems.
In fact, we find stable asymmetric configurations in the range of $250 \leq \sedimentationStrength \, d^4 / k_B T \leq 450$.
For smaller values of $\sedimentationStrength$, the sedimentation force is insufficient to prevent mixing.
On the opposite side, larger values of $\sedimentationStrength$ lead to reentrant mixing due to an unrealistically strong compression of the layers.

We consider $N=1058$ and $N = 1587$ large particles for the one-component bilayer and trilayer system, respectively.
The two-component system consists of $N_1 = 1058$ large particles and $N_2=529$ small particles.
All systems were equilibrated for more than $10^7$ steps ($t > 100\tau$), with the discrete time step $\delta t = 10^{-5}\tau$.
After that, the shear force was switched on and the simulation was carried out for an additional time period of $t = 100\tau$, in which the systems reached a steady state.
Only after this period we started with the calculation of material properties.
\section{\label{sec:average motion}Simulation results for average motion}
\begin{figure*}
  \includegraphics[width=1.0\linewidth,natheight=1168,natwidth=4260]{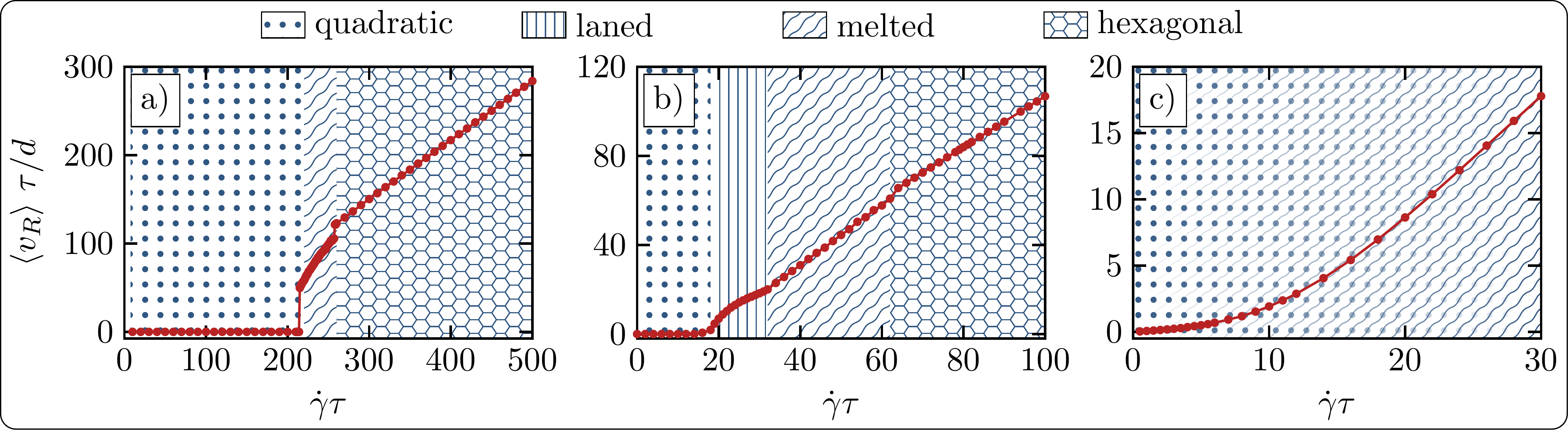}
  \caption{
    (Color online) Average velocity $\av{v_R} $ in flow ($x$-) direction of the top layer relative to the bottom layer(s) for a) the one-component bilayer, b) the one-component trilayer and c) the two-component trilayer system. The corresponding in-plane structure is indicated by the filling pattern.
  }
  \label{fig:average motion}
\end{figure*}
As a starting point, we analyze the dynamics of the model systems by calculating the average velocity $\av{v_R}$ of the crystal layers in flow ($x$-) direction relative to each other.
The average relative velocity in $y$- and $z$-direction vanishes for all considered systems.
Results for $\av{v_R}$ in the one-component bilayer and trilayer system as well as the two-component system are plotted in Figs. \ref{fig:average motion}a)-c).
In those figures, the dynamical states of the considered systems are indicated by different patterns.
These states were distinguished by monitoring the four- and sixfold in-plane angular bond order parameters $\Psi_4, \Psi_6$ \cite{Vezirov2013}.

For the one-component bilayer system [\fref{average motion}a)], we observe a pronounced depinning transition at the critical shear rate $\shearrate_c^{\text{BD}}\tau \approx 214$.
For subcritical shear rates $\shearrate < \shearrate_c^{\text{BD}}$, the system is "locked", with the colloids being pinned (apart from thermal fluctuations) on the sites of the crystalline layers with quadratic in-plane structure.
Increasing the shear rate then leads to a depinning of the crystalline layers and melting of the in-plane structure.
For large shear rates, a hexagonal crystalline order is recovered, which is accompanied by a collective zig-zag motion of the colloidal crystal layers \cite{Vezirov2013}.
A similar depinning transition (yet no subsequent crystallization) is found for driven monolayers on a periodic potential \cite{Achim2009,Bohlein2012,Hasnain2013,Juniper2015}.
Using this connection, we can formulate a simple model to estimate the critical shear rate.
This is discussed in detail in \sref{bilayer}.

We now consider the one-component trilayer system.
Here, the dynamics can be characterized by the average velocity of the top layer relative to the two bottom layers, which is plotted in \fref{average motion}b) \footnote{
\label{note: symmetric trilayer}
  For the symmetric trilayer system, the middle layer does not move $\av{v_{mid} } = 0$ and the outer layers move with the same velocity $\av{v_{out} } $ in opposite directions, consistent with the velocity profile imposed by the linear shear flow. Therefore, the velocity of the top layer relative to the two bottom layers is given by $ \av{v_R} = \av{v_{out} } - ( \av{v_{mid} }  - \av{v_{out} } )/2 = 1.5 \av{v_{out} } $
}.
In contrast to the bilayer system, the trilayer system displays a continuous onset of motion (i.e., no jump of the velocity) due to a new intermediate laned state.
Again, for small shear rates the colloidal layers are pinned in quadratic in-plane lattices.
However, upon increasing the shear rate, the middle layer becomes unstable and splits into two sublayers, which are each pinned to one outer layer.
The colloids in the sublayers form lanes, moving with the velocity of the respectively closest outer layer \cite{Vezirov2015a}.
This leads to a nonlinear velocity profile $\av{v_R}\lr{z}$ until the melted state is reached, where a quasi-linear velocity profile is recovered.
For large shear rates, the system forms a hexagonal steady state, similar to the bilayer system (see also \sref{trilayer}).

Introducing a second species to the system, the average velocity behaves very different to the one-component systems, as seen in \fref{average motion}c).
Specifically, we consider the average velocity of the top layer consisting of small colloids relative to the bottom layers consisting of large colloids.
The latter are locked in a quadratic crystalline structure for all considered shear rates.
Contrary to the one-component systems, the top layer of the binary system is never pinned to the bottom layers.
Instead, the top layer (which is weakly ordered, i.e., $\Psi_4 = 0.7$ and $\Psi_6=0.36$ for $\shearrate\tau=0$) transitions continuously into a melted state with increasing shear rate.
This is accompanied by a continuous onset of motion and results in a finite average velocity for all nonzero shear rates.
In order to understand this dynamics, we investigate the local structure and dynamics of the top layer in \sref{density excitations}.
\section{\label{sec:bilayer}Theory of depinning in the bilayer system}
In this section we will present a simple model, which allows us to estimate the critical shear rate of the depinning transition based on the equilibrium configuration.
Within this model, we map the dynamics of the bilayer system to the motion of a single particle in a 1D periodic potential.
This is in the spirit of the FK model \cite{Braun2000}, which considers a 1D chain of (harmonically) coupled colloids on a periodic sinusoidal substrate potential.
Importantly, the resulting equation of motion can be solved analytically, allowing a direct (yet approximate) calculation of the average relative velocity and also of the shear stress of the bilayer system.
\subsection{\label{sec:driven monolayers}Driven monolayers}
The 1D overdamped equation of motion for a particle $i$ in a driven monolayer is given by \cite{Hasnain2013, Risken1996}
\begin{equation}\label{eq:equation of motion driven monolayer}
	\mu^{-1} \dot{x}_i = \sum_{j \neq i}^{N_L} F_{inter}\lr{x_{ij} } + F_{sub}\lr{x_i} + F_d + \Gamma_{i}\lr{t}
	\text{,}
\end{equation}
with $N_L$ being the number of particles in the monolayer, $F_{inter}$ the two-particle interaction force, $F_{sub}$ the periodic substrate force, $F_d$ the constant driving force, and $\Gamma_{i}=\mu^{-1}\dot{W}_i$ the random force.

In the following we focus on a special case, which involves an infinitely stiff crystalline monolayer (corresponding to the strong coupling limit).
In this limit, the average velocity of all particles is determined by the velocity of the center of mass, $X$, where $X = N_L^{-1}\sum_{i=1}^{N_L} x_i$.
Indeed, for a large number of particles, $N_L \to \infty$, the random forces acting on $X$ vanish.
Further, considering radial pair interactions, the sum of all interaction forces vanishes due to the crystal symmetries.
We can thus restrict our consideration to the motion of $X$, determined by
\begin{equation}\label{eq:equation of motion stiff monolayer}
	\mu^{-1} \dot{X} = F_{sub}\lr{X} + F_d
	\text{.}
\end{equation}
For a sinusoidal substrate force $F_{sub}\lr{X} = F_{max} \sin\lr{ 2\pi X / a }$, this equation can be solved analytically \cite{Risken1996}.
The resulting average relative velocity is given by \cite{Hasnain2013}
\begin{equation}\label{eq:mean velocity stiff monolayer}
	\av{v_R}  = a \lr{\int_{0}^{a} \dot{X}^{-1} dX }^{-1}	= \mu\sqrt{F_d^2-F_{max}^2}
	\text{.}
\end{equation}
Equation (\ref{eq:mean velocity stiff monolayer}) expresses the fact that the crystal monolayer is pinned ($\av{v_R} = 0$) for driving forces smaller than the critical driving force ($F_{d,c} = F_{max}$) and displays a running state ($\av{v_R} > 0$) for larger driving forces.
\subsection{\label{sec:mapping to shear-driven system}Mapping to shear-driven system}
In order to relate the behavior of the driven monolayer to the dynamics of colloidal layers under shear flow, we need to formulate, for the shear-driven systems, an effective substrate force as well as an effective driving force.
To this end, we focus on the dynamics of the top layer, whereas the bottom layer is assumed to act as a "substrate".
From \eref{equation of motion}, the equation of motion of the center of mass $\Delta\vec{R} = N_L^{-1} \sum_{i=1}^{N_L} \position_i - \vec{R}_{bot}$ of the top layer relative to the center of mass of the bottom layer ($\vec{R}_{bot}$) in flow ($x$-) direction follows as
\begin{widetext}
\begin{equation}\label{eq:equation of motion center of mass}
	\Delta\dot{R}_x = \frac{\mu}{N_L} \sum_{i=1}^{N_L} \lr{ \sum_{j\neq i}^N F_{x}^{inter}\lr{ r_{ij} } + F_x^{wall} + \mu^{-1}\dot{W}_{i,x}  }+ \shearrate \Delta R_z \text{,}
\end{equation}
\end{widetext}
where $N_L$ is the number of particles of the top layer.
Again, for $N_L \to \infty$, the mean of the random forces acting on the layer vanishes, i.e., $N_L^{-1} \sum_{i=1}^{N_L} \mu^{-1}\dot{W}_{i,x} \approx 0$.
The force exerted from the confinement (see \eref{soft-wall potential}) has no $x$-component, thus $F_x^{wall}=0$.
Comparing the remaining terms with \eref{equation of motion stiff monolayer}, we identify the shear force as the driving force, i.e.,
\begin{equation}\label{eq:driving force}
F_d\lr{\Delta R_z}=\mu^{-1} \shearrate \Delta R_z \text{.}
\end{equation}
Further, the sum of particle interaction forces acting on the layer can be identified as the substrate force, i.e.,
\begin{equation}\label{eq:substrate force}
 F_{sub}\lr{\llrr{ \position } } = \frac{1}{N_L}\sum_{i=1}^{N_L} \sum_{j\neq i}^N F_{x}^{inter}\lr{ r_{ij} } \text{,}
\end{equation}
where $\force^{inter} = -\nabla_{\position_i} U_{inter}$ is the particle interaction force [see \eref{total particle interaction}] and $\llrr{ \position } = \position_1 ,\ldots, \position_N $ is the set of particle positions.
Here we are interested in the depinning starting from the quadratic (equilibrium) state.
The particle positions (in the absence of noise) are therefore given by $ \position \lr{t} = \position_{nm} + \vec{R}\lr{t} $, with the corresponding lattice position $\position_{nm}$ and the center of mass of the layer, $\vec{R}$.
In this framework, the position on the lattice is given by the primitive vectors and is constant.
Using this ansatz, we can rewrite the substrate force (\ref{eq:substrate force}) for particles of the top layer
\begin{align}\label{eq:reduced substrate force}
	F_{sub}\lr{\Delta \vec{R} } &= F_{x}^{inter,bot}\lr{ \Delta \vec{R} } \nonumber\\
	 &= \frac{1}{N_L}\sum_{i=1}^{N_L} \sum_{j = 1}^{N_{bot}} F_{x}^{inter}\lr{ r_{ij} } \text{,}
\end{align}
where $\force^{inter,bot}$ is the sum of interaction forces between particles of the top layer and particles of the bottom layer and $N_{bot}$ is the number of particles in the bottom layer.
The corresponding sum within the top layer is zero due to the crystal symmetry.

Inserting \eref{driving force} and \eref{reduced substrate force} into \eref{equation of motion center of mass} and neglecting the noise we obtain
\begin{equation}\label{eq:reduced equation of motion center of mass}
		\mu^{-1}\Delta \dot{R}_x = F_{sub} \lr{\Delta \vec{R} } + F_d \lr{\Delta R_z}\text{.}
\end{equation}
The structure of \eref{reduced equation of motion center of mass} is already close to the corresponding monolayer equation \eref{equation of motion stiff monolayer}, yielding the strategy to calculate the critical shear rate via \eref{mean velocity stiff monolayer}.
However, in \eref{reduced equation of motion center of mass}, both the driving force (\ref{eq:driving force}) and the substrate force (\ref{eq:reduced substrate force}) still depend on the layer distance $\Delta \vec{R}$.
To proceed, we make the following \emph{ansatz} for $\Delta \vec{R}$ as function of the (relative) displacement of the center of mass,
\begin{equation}\label{eq:layer distance}
\Delta \vec{R}\lr{X} = X \vec{e}_x + \Delta R_{y}^{eq} \vec{e}_y + \Delta R_{z}\lr{X} \vec{e}_z\text{.}
\end{equation}
According to \eref{layer distance} we set the $x$-component of $\Delta \vec{R}$, $\Delta R_x$, equal to the variable $X$, which represents the center-of-mass coordinate in the 1D driven monolayer [see \eref{equation of motion stiff monolayer}].
Further, the $y$-component is set to its equilibrium value, which is constant ($\Delta R_y^{eq}$).
Indeed, from the symmetry of the system, it follows that $\Delta\dot{R}_y = F_y^{inter,bot}\lr{ \Delta R_y^{eq} } = 0$.
However, this does not hold for the displacement in $z$-direction $\Delta R_z$.
\begin{figure}
	\includegraphics[width=1.0\linewidth]{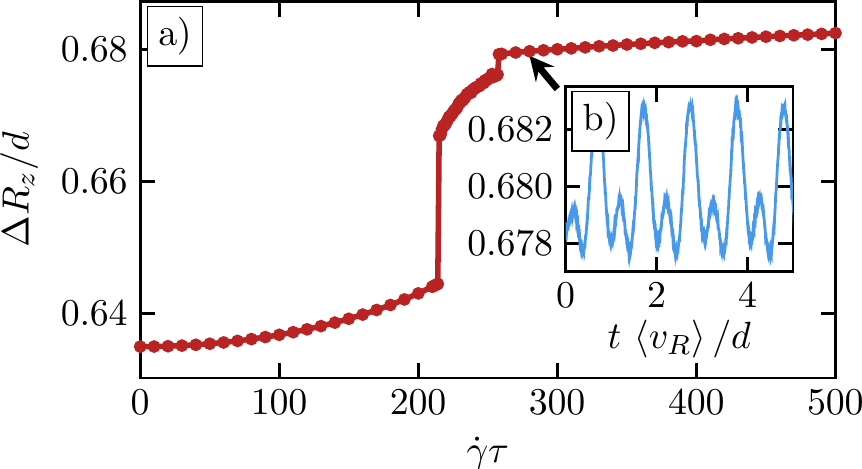}
	\caption{(Color online) a) Mean layer distance of the one-component bilayer system in dependence of the shear rate. b) Time dependence of the layer distance for $\shearrate\tau=280$ in the hexagonal steady state.}
	\label{fig:mean layer distance bilayer}

\end{figure}
In fact, $\Delta R_z$ depends markedly on the shear rate, see \fref{mean layer distance bilayer}.
In particular, one observes a pronounced increase of $\Delta R_z$ when the system transforms from the quadratic into the hexagonal phase.
Moreover, within the hexagonal state, $\Delta R_z$ actually oscillates in time, mimicking the zig-zag motion of the particles \cite{Vezirov2013}.

In view of the strong dependence of $\Delta R_z$ on the shear rate, it is not surprising that setting $\Delta R_z$ to its constant equilibrium value ($\shearrate\tau=0$) and using this value for the calculation of $F_d$ and $F_{sub}$ yields a wrong result (specifically an overestimation) for the critical shear rate.
Indeed, this simple calculation yields $\shearrate_c\tau \approx 330$, which has to be compared to the true value (obtained from BD simulation) of $\shearrate_c^{\text{BD}}\tau \approx 214$.
A somewhat better result is obtained if one sets $\Delta R_z = \Delta R_z\lr{\shearrate}$.
However, this requires to compute the nonequilibrium properties of the considered system beforehand.
A more desirable strategy would be to define all ingredients for the calculation of the critical shear rate based on the \emph{equilibrium} configuration.
To this end we model the motion of the particles by an optimal path defined for the \emph{equilibrium} configuration (see \eref{optimal path} in Appendix \ref{sec:appendix:optimal path}).
This allows to obtain analytic expressions for $F_{sub}$ and $F_d$.

Inserting \eref{effective substrate force} and (\ref{eq:layer distance optimal path}) from Appendix \ref{sec:appendix:optimal path} into \eref{reduced equation of motion center of mass}, we obtain an equation for the relative motion of the layers in $x$-direction
\begin{equation}\label{eq:equation of motion simple model}
	\dot{X} = \mu \, F_{max} \sin\lr{ \frac{2\pi}{a} X } + \shearrate Z_A \cos\lr{ \frac{2\pi}{a} X } + \shearrate Z_0 ,
\end{equation}
This equation can be solved analytically, the resulting trajectories $X\lr{t}$ are given in \eref{trajectories optimal path} in Appendix \ref{sec:appendix:trajectory}.
The average velocity of the layers then follows as
\begin{equation}\label{eq:relative velocity for optimal path}
	\av{v_R} = \sqrt{ \shearrate^2 \lr{ Z_0^2 - Z_{A}^2 } - \mu^2 F_{max}^2 }\text{,}
\end{equation}
yielding the critical shear rate
\begin{equation}\label{eq:critical shear rate}
	\shearrate_c = \frac{\mu\,F_{max} }{ \sqrt{Z_0^2-Z_A^2} } \text{.}
\end{equation}
From the trajectories $X\lr{t}$, particularly their long-time solution $\tilde{X}\lr{t}$ given in \eref{long-time position} in Appendix \ref{sec:appendix:trajectory}, we can further calculate the mean shear stress.
The latter is determined (neglecting kinetic contributions \cite{Vezirov2015}) via the $x$-$z$-component of the stress tensor,
\begin{equation}\label{eq:shear stress}
	\sigma_{xz} = \av{ \frac{1}{V} \sum_{i}\sum_{j>i} F_{x}^{inter}\lr{r_{ij}}z_{ij} }\text{,}
\end{equation}
where $V$ is the volume of the simulation box and $z_{ij}$ is the particle distance in $z$ direction.
Within our simple model, the shear stress for particles of the same layer vanishes ($z_{ij} = 0$).
Therefore, the relevant particle interaction forces are given by the substrate force $F_{sub}$ [see \eref{effective substrate force} in Appendix \ref{sec:appendix:optimal path}] and the particle distance is defined by the layer distance $\Delta R_z\lr{X}$ [see \eref{layer distance optimal path}].
The mean shear stress of the system then reads
\begin{equation}\label{eq:shear stress simple model}
	\sigma_{xz} = \frac{N}{4Vt_{a}} \int_{0}^{t_{a}} F_{sub}\lr{ \tilde{X}\lr{t} }\Delta R_{z}\lr{ \tilde{X}\lr{t}}dt\text{,}
\end{equation}
where $N$ is the number of particles and $t_{a}=a/\av{v_R}$ is the time period for the top layer to move over one lattice position.
\subsection{\label{sec:bilayer results}Numerical results for the bilayer system}
\begin{figure}
	\includegraphics[width=1.0\linewidth]{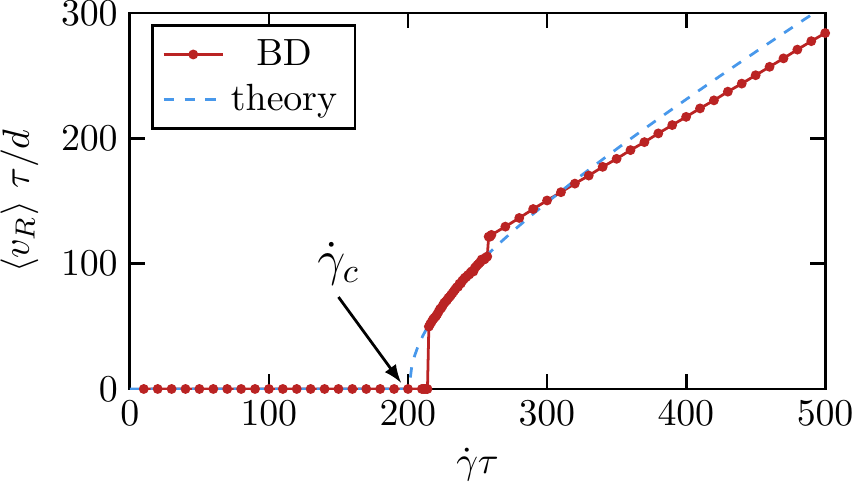}
	\caption{(Color online) Average velocity of the top layer relative to the bottom layer in the bilayer system from BD simulations (red) and from \eref{relative velocity for optimal path} (blue dashed), revealing the critical shear rate $\shearrate_c$.}
	\label{fig:relativeVelocityBilayer}
\end{figure}
To judge the performance of the effective theory, outlined in \sref{mapping to shear-driven system}, we compare in \fref{relativeVelocityBilayer} the average velocity numerically obtained from \eref{relative velocity for optimal path} with corresponding BD simulation data.
Focusing first on the critical shear rate $\shearrate_c$, we find that the effective model is in good quantitative agreement ($\shearrate_c\tau\approx200$) with the BD results ($\shearrate_c^{\text{BD}}\tau\approx214$).
However by construction, the model predicts a \emph{continuous} transition from the pinned to the free sliding state.
This is clearly in contrast to the BD results, which indicate a discontinuous transition (accompanied by jumps in the velocity) from the quadratic to the melted state, as well as from the melted to the hexagonal state.
As analyzed in Ref. \cite{Achim2009}, these discontinuous transitions are related to the shear-induced restructuring of the in-plane order [see also \fref{average motion}a)].
Obviously, these complex processes are beyond the scope of the proposed model.
Still, the good estimate for $\shearrate_c$ suggests that the impact of structural changes occuring at larger $\shearrate$, as well as of thermal noise can be neglected if we just focus on the depinning itself.
A further interesting aspect arising from the effective model is that the critical shear rate [see \eref{critical shear rate}] strongly depends on the distance of the layers in $z$-direction.
In particular, an increase of the layer distance leads to the reduction of the critical shear rate (due to the increase of $F_d$ as well as the decrease of $F_{max}$).
This is a physically plausible result.\\
\begin{figure}
	\includegraphics[width=1.0\linewidth]{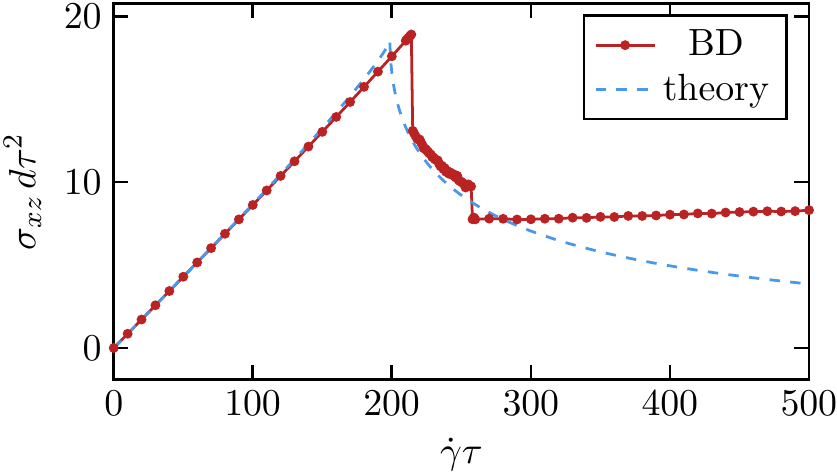}
	\caption{(Color online) Shear stress of the bilayer system from BD simulations (red) and from \eref{shear stress simple model} (blue dashed).}
	\label{fig:bilayer rheology}
\end{figure}
We now turn to the shear stress, $\sigma_{xz}$, in the long-time limit.
In \fref{bilayer rheology} we compare the shear rate dependence of $\sigma_{xz}$ obtained from \eref{shear stress simple model} with corresponding BD results \cite{Vezirov2015}.
Starting from the equilibrium configuration, the simulated system displays a quasi-linear increase of $\sigma_{xz}$, corresponding to an elastic deformation of the quadratic structure in the crystalline layers.
Once the system melts, the system becomes mechanically unstable, as reflected by the negative slope in $\sigma_{xz}$.
Finally, after the recrystallization into a hexagonal lattice $\sigma_{xz}$ increases again with $\shearrate$ \cite{Vezirov2015}.
Similar to these BD results, the effective model predicts an approximately linear increase of $\sigma_{xz}$ for subcritical shear rates $\shearrate \leq \shearrate_c$ as well as a sharp, nonlinear increase close to the critical shear rate.
For supercritical shear rates $\shearrate > \shearrate_c$, the shear stress then decreases essentially exponentially (as seen from a logarithmic plot) to zero, which corresponds to the shear stress of a freely sliding layer.

Overall, the effective model thus provides a reasonable description of the shear stress within the quadratic and melted state, similar to the estimated average velocity discussed before.
However, for $\shearrate \gg \shearrate_c$, the shear stress deviates markedly from that of the true system, where the structure becomes again crystalline and the particles perform a characteristic a zig-zag motion in $y$-direction \cite{Vezirov2013}.
\section{\label{sec:trilayer}Depinning in the symmetric trilayer system}
\begin{figure}
	\includegraphics[width=1.0\linewidth]{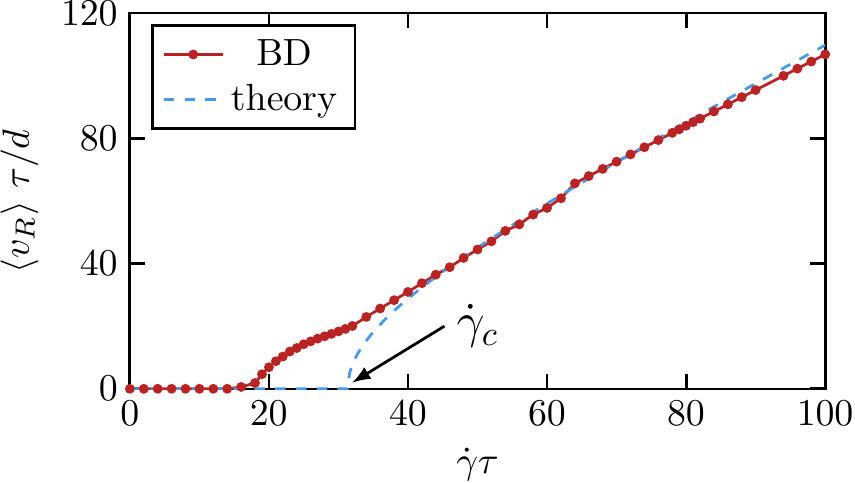}
	\caption{(Color online) Average velocity of the top layer relative to the two bottom layers of the symmetric trilayer system from BD simulations (red) and the simple model \eref{relative velocity for optimal path} (blue) with critical shear rate $\shearrate_c$.}
	\label{fig:relative velocity trilayer optimal}
\end{figure}
As discussed in \sref{average motion}, the one-component trilayer system also displays a depinning transition similar to the bilayer system (see also Ref. \cite{Vezirov2015}).
Applying the mapping strategy presented in the previous section to the trilayer system, we can calculate the mean relative velocity of the top layer relative to the two bottom layers \cite{Note1}, which is plotted in \fref{relative velocity trilayer optimal}.
Contrary to the case of the bilayer, we find that the model here overestimates the critical shear rate.
This is due to the additional laned state \cite{Vezirov2015a}, in which the middle layer becomes unstable and splits into two sublayers.
Still, closer inspection shows that the model does predict the onset of the melted state (which occurs at $\shearrate\tau \approx 34$ according to the BD simulation) in good quantitative agreement with BD data.
This suggests that the melting of the crystal layers is indeed induced by the depinning of the outer layers.
For even larger shear rates, the average velocity of the simple model is, in fact, in nearly perfect agreement with the corresponding BD result, despite the fact that the true system has undergone an additional structural transition from a melted into a hexagonal state [see \fref{average motion}c)].
\section{\label{sec:density excitations}Asymmetric Trilayer: Density excitations}
\begin{figure}
  \includegraphics[width=1.0\linewidth,natheight=746,natwidth=2027]{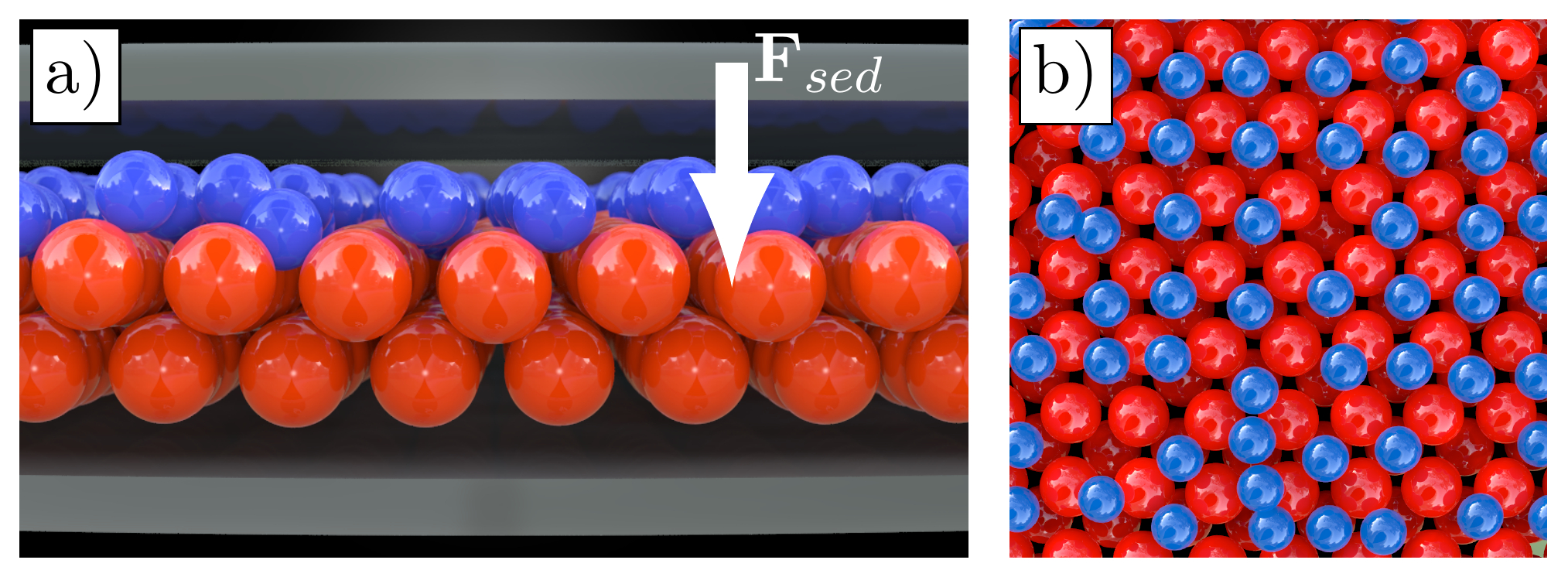}
	\caption{(Color online) a) Side view and b) top view on the binary system in equilibrium ($\shearrate\tau=0$), displaying two quadratic bottom layers (red) and one top layer containing small particles (blue). }
	\label{fig:binary configuration}
\end{figure}
In this section we turn to a binary system of large and small colloids, where the different sizes induce a mismatch of the structural length scales of the corresponding pure systems.
Applying a constant sedimentation force $F_{sed}=-\nabla_{\position}U_{sed}\lr{z_i}$ [see \eref{sedimentation potential}], we can stabilize asymmetric configurations already at $\shearrate\tau = 0$.
These consist of two bottom layers containing only large colloids and one layer of small colloids on top, as shown in \fref{binary configuration}a).
The large colloids form crystalline layers with quadratic in-plane structure.
This structure, in turn, induces a semi-crystalline structure (characterized by order parameter values $\Psi_4 = 0.7$ and $\Psi_6=0.36$ at $\shearrate\tau=0$) of the particles in the top layer [see \fref{binary configuration}b)].
We note that the density of small particles is chosen such that, in principle, all "potential valleys" created by the bottom layers are filled with exactly one small particle.
For this density, the equilibrium structure of the small particles alone is liquidlike.

For the following investigations under shear, we will consider only shear rates which are subcritical with respect to the depinning of the two bottom layers, as well as insufficient to introduce a mixing of the two colloidal species.
The critical shear rate of the bottom layers follows from \eref{critical shear rate} as $\shearrate_c\tau \approx 98$.
We note that the range of relevant shear rates depends on the sedimentation potential strength $\sedimentationStrength$, since the latter influences the layer distance.
In contrast, we find that the dynamical behavior (in particular, the relation between the average- and the kink velocity to be discussed in \sref{estimate average motion via local dynamics}) is  rather independent of the particular choice of $\sedimentationStrength$.
\subsection{\label{sec:structural properties}Structural properties of the top layer}
\begin{figure}
	\includegraphics[width=1.0\linewidth]{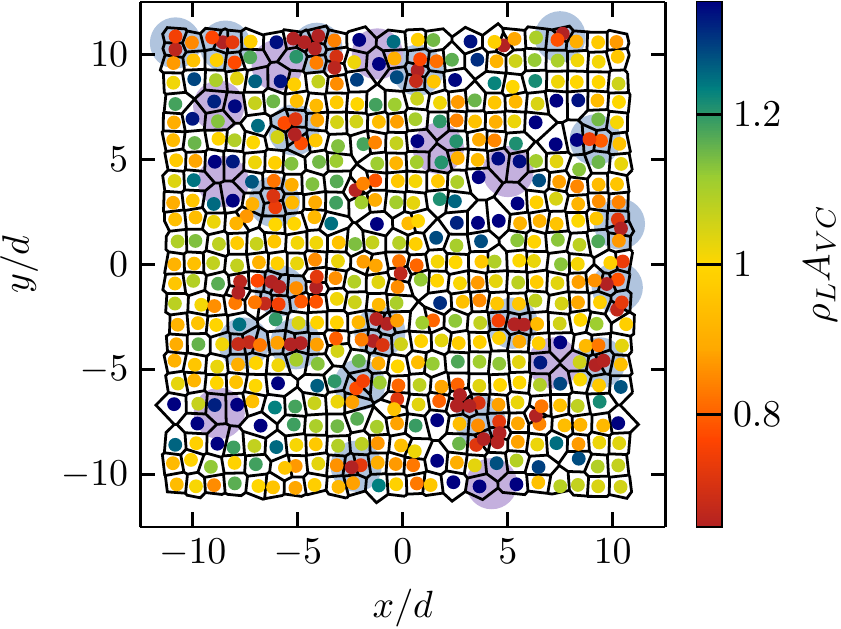}
	\caption{(Color online) Top view on the equilibrium configuration ($\shearrate\tau=0$) and corresponding Voronoi tessellation (black) of particles of the top layer. The particles are colored with respect to their normalized Voronoi cell areas $\rho_L A_{VC}$. The position of antikinks $\rho_L A_{VC} > 1.2$ (violet) and kinks $\rho_L A_{VC} < 0.8$ (gray) are determined via a cluster identification algorithm.}
	\label{fig:voronoi configuration}
\end{figure}
To analyze the local structure of the top layer we calculate the corresponding 2D Voronoi tessellation \cite{Aurenhammer1991}, which divides the total area of the layer into "eigencells".
Each eigencell contains exactly one particle.
The boundaries of each eigencell follow from analyzing the connecting vectors $\vec{r}_{ij}$ of the central particle with all of its neighbors; each boundary then corresponds to the perpendicular bisector of $\vec{r}_{ij}$, i.e., a line perpendicular to $\vec{r}_{ij}$ and cutting $\vec{r}_{ij}$ at its half.
The resulting area of the Voronoi cell, $A_{VC}$, allows to define a local density proportional to the inverse $A_{VC}$.
The Voronoi tessellation of the top layer in equilibrium ($\shearrate\tau=0$) is shown in \fref{voronoi configuration}.
In this figure, the particles are colored according to their normalized Voronoi cell area $\rho_L A_{VC}$, where $\rho_L = N_L / L^2$ is the average 2D number density of the layer ($L$ is the length of the simulation box and $N_L$ is the number of particles in the top layer).
In a \emph{perfect} lattice, one would have $\rho_L A_{VC}=1$ throughout the system.
Inspecting \fref{voronoi configuration}, we find that the true structure in the top layer is characterized by a substantial amount of defects.
Specifically, one observes both, cells with enhanced area relative to the ideal case (corresponding to a smaller-than-average local density) and cells with reduced area (corresponding to a locally increased density).
In analogy to the 1D FK model we call these defects "antikinks" ($\rho_L A_{VC} > 1.2$) and "kink" ($\rho_L A_{VC} < 0.8$) \cite{Braun1998, Braun2000}, respectively.
In the original FK model, an ideal kink consists of a single additional particle on a fully occupied lattice \cite{Braun1998}.
This additional particle can push another particles to the next occupied lattice position, leading to a hopping wave.
Similarly, an ideal antikink corresponds to a missing particle, allowing the neighboring particles to pull a particle to the unoccupied lattice side.
In other words, the kinks (antikinks) imply that there is more than (less than) one particle per lattice side.
Contrary to that, we find that in our system most of the defects are formed by multiple additional or missing particles.
Furthermore the defects extend over several lattice sides.

\begin{figure}
	\includegraphics[width=1.0\linewidth]{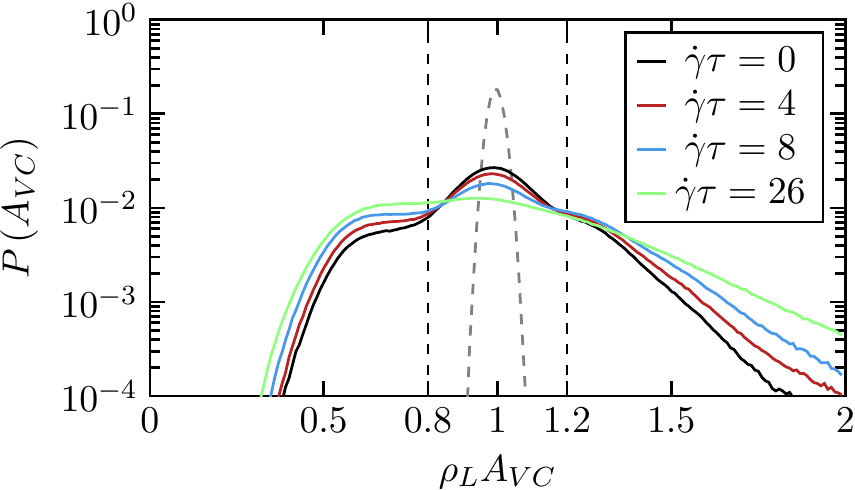}
	\caption{(Color online) Distribution of the Voronoi cell areas in the top layer for different shear rates $\shearrate\tau=0,4,8,26$ (black, red, blue, green respectively). The Voronoi cell area distribution of the quadratic bottom layers (gray dahsed) is plotted as a reference. The vertical dashed lines indicate the threshold values for kinks ($\rho_L A_{VC} < 0.8$) and antikinks ($\rho_L A_{VC}>1.2$).}
	\label{fig:voronoi cell area histogram}
\end{figure}
To quantify the number of particles contributing to defect structures we have calculated the time averaged distribution of Voronoi cell areas, which is plotted in \fref{voronoi cell area histogram}.
Included is the result for the bottom layers (dashed line).
These form a nearly perfect quadratic structure as reflected by the sharp peak at $\rho_L A_{VC}=1$.
Inspecting now the top layer distribution we observe, at $\shearrate\tau=0$, that $P\lr{A_{VC}}$ still has a maximum at $\rho_L A_{VC} = 1$.
However, there are also pronounced, asymmetric flanks, corresponding to particles in kinks ($\rho_L A_{VC} < 0.8$) and antikinks ($\rho_L A_{VC} > 1.2$).
We note that the left-hand flank is bounded by the tightest possible packing of small colloids.
This explains the rapid decrease of the number of particles with $\rho_L A_{VC} < 0.5$.
Such a limitation does not exist for the number of antikinks, which explains the much broader shape of $P\lr{A_{VC}}$ at the right side.

Considering now the impact of shear, we observe, first, a progressive decrease and finally, a disappearance, of the maximum of $P\lr{A_{VC}}$ at $\rho_L A_{VC}=1$.
This reflects the decrease of the number of particles with local quadratic order.
At the same time, the number of particles involved in kinks and antikinks increases.
Specifically, we observe that the area distribution for kinks increases mainly in height, but not in width, consistent with the above-mentioned limitation.
Therefore, the number of kinks with similar values of the local density increases with $\shearrate$.
This is in contrast to the antikinks, whose area distribution increases mainly in width, corresponding to an increasing size of defect structures with multiple missing particles.
Finally, for shear rates beyond the critical shear rate ($\shearrate_c\tau\approx 21$) of the idealized (crystalline) top layer (given by \eref{critical shear rate}), the semi-quadratic structure of the real top layer is essentially lost and most particles contribute to large defect structures.
\subsection{\label{sec:single particle and cluster dynamics}Single particle and cluster dynamics}
\begin{figure}
	\includegraphics[width=1.0\linewidth]{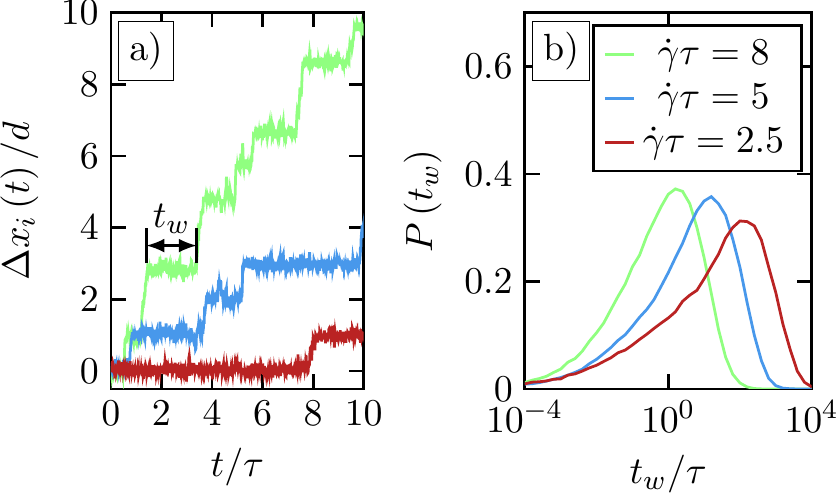}
	\caption{(Color online) a) Displacement $\Delta x_i\lr{t} = x_i\lr{t}-x_i\lr{0}$ ($x$-direction) of a randomly chosen single particle in the top layer as function of dimensionless time and b) distribution of the waiting times $t_w$ for various shear rates.}
	\label{fig:single particle trajectories}
\end{figure}
We now turn to the time-resolved dynamical behavior.
To start with, we plot in \fref{single particle trajectories}a) the displacement of a single particle in the top layer in $x$-direction for different shear rates.
In all cases the particle spends relatively long time at a lattice position before jumping to the next one.
In other words, the waiting time $t_w$ (defined according to the "minimum-based" definition in \cite{Gernert2014}) is larger than the Brownian timescale $\tau$ characterizing the diffusion of the free small particle over the distance $d$.
On increasing $\shearrate$, the jumps become more frequent, as expected due to the stronger drive which helps to overcome the "barriers" generated by the bottom layers.
This is also reflected by the distribution of the waiting times [see \fref{single particle trajectories}b)], whose maximum shifts to shorter times for increasing shear rates.
At this point we recall the increase of the number of kinks with $\shearrate$ discussed in \sref{structural properties}.
Having this in mind, the enhancement of the jump frequency (i.e., $1/t_w$) may be taken as an indication that the jumping particle is part of a kink.
We also note that, in contrast to the FK model, the particles in the present system can jump multiple lattice sides at once.
Clearly (see \fref{single particle trajectories}) this becomes more likely for large shear rates (e.g. $\shearrate\tau = 8$).

\begin{figure}
	\includegraphics[width=1.0\linewidth]{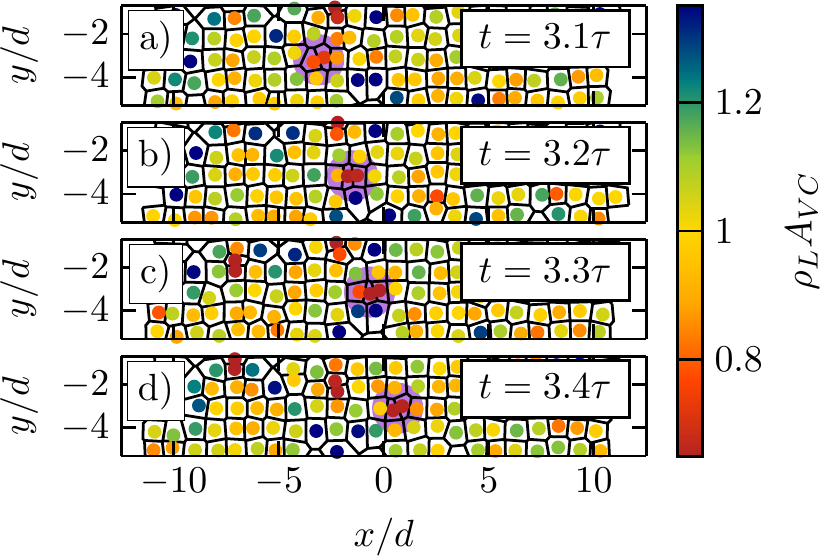}
	\caption{(Color online) Example of the motion of density excitations in the top layer. Parts a)-d) show a section of the top layer (color coded Voronoi tessellation) at four different time steps. The circle (purple) indicates the time-dependent position of a kink ($\rho_L A_{VC} < 0.8$).}
	\label{fig:voronoi configuration dynamics}
\end{figure}
In addition to tracking single particles, we have also investigated the motion of defect structures (kinks and antikinks) involving several particles.
An example of this analysis is shown in \fref{voronoi configuration dynamics}, where a section of the top layer is plotted at four different times.
The series clearly reveals the motion of a kink in $x$-direction.
This kink is tracked via a modified Hoshen-Kopelman algorithm, which identifies clusters with enhanced local density (i.e., $\rho_L A_{VC} < 0.8$) on the underlying triangular lattice given by the Delaunay triangulation.
For antikinks, the same approach is used to track clusters with reduced local density (i.e., $\rho_L A_{VC} > 1.2$).

\begin{figure}
	\includegraphics[width=1.0\linewidth]{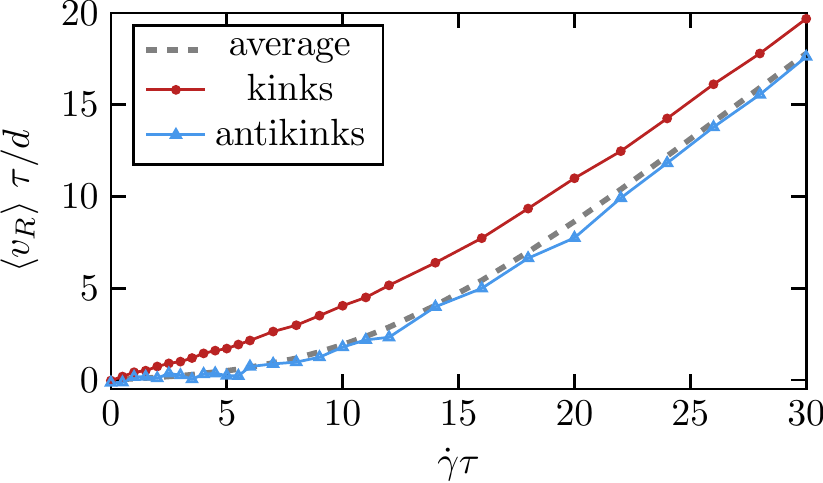}
	\caption{(Color online) Velocity of the kinklike and antikinklike defects relative to the bottom layers as functions of the shear rate. The relative velocity of the top layer (gray, dashed) is plotted as a reference.}
	\label{fig:relative velocity density excitations}
\end{figure}
The tracking of the positions of the kinks and antikinks furthermore allows us to calculate their average velocity relative to the bottom layers in $x$-direction.
The resulting velocities are shown in \fref{relative velocity density excitations} as functions of the shear rate.
For comparision, we have included the average relative velocity of the top layer.
In equilibrium (i.e., $\shearrate\tau=0$), the kinks display no net motion just like the top layer.
This picture changes at finite shear forces, where the kinks move \emph{faster} than the average (i.e., $\av{v_{kink}}>\av{v_R}$).
This holds for all shear rates considered, however, the difference (more specifically, the ratio $\av{v_{kink}}/\av{v_R}$) is largest in the range $0<\shearrate\tau<5$.
Here the velocity of the kinks is nearly one order of magnitude larger than the velocity of the top layer.
This observation is in accordance with a prediction from the FK model, where the monolayer is displaced exactly one lattice side when a single kink travels through the layer \cite{Braun1998}, i.e.,
\begin{equation}\label{eq:FK velocity relation}
  \av{v_R} = \frac{N_K}{N_L} \av{v_{kink} } \text{,}
\end{equation}
with $N_K$ the number of kinks and $N_L$ the number of particles in the monolayer.
Therefore, if there is only a small number of kinks, the velocity of the layer is expected to be much slower than the velocity of the kinks.
We will come back to this point in the subsequent section~\ref{sec:estimate average motion via local dynamics}.
Increasing the shear rate leads to a corresponding increase of the number of kinks (see \fref{voronoi cell area histogram}).
As a consequence the difference between the velocities decreases.

In contrast to the kinks, the antikinks seem to be "locked" within the top layer as revealed both by the direct visualization in \fref{voronoi configuration dynamics} and by the fact that their average velocity (see \fref{relative velocity density excitations}) is nearly identical to that of the top layer.
This "locking" behavior of the antikinks is in contrast to the (original) FK model, where the antikinks move with a velocity which is the same in magnitude, but opposite in direction to that of the kinks.
The reason for the antikink motion in the FK model is the attractive harmonic interaction potential linking the particles \cite{Braun1998}.
In driven monolayers with purely repulsive interactions, the magnitude of the velocity of the antikinks is expected to be smaller than that of the kinks \cite{Bohlein2012, Hasnain2013} but still different from the average motion of the layer.

In our system, the antikinks apparently move \emph{along} the direction of the driving force, which can not be explained by the absence of attractive interactions alone.
Instead, we interpret this phenomena as a result of the the fact that, in our system, the "substrate" acts not as an \emph{external} potential, but as a part of the layered system which responds to the behavior of the top layer.
Indeed, we find that the large particles of the bottom layer shift to higher $z$-positions in the vicinity of the antikinks.
In other words, the reduced local density in the top layer leads to a bump formed by the bottom particles.
These deformations of the bottom layers (which correspond to higher potential barriers) then prevent particles of the top layer to jump into the empty lattice positions.
Instead, the antikinks are pushed in the direction of the driving force.
\subsection{\label{sec:estimate average motion via local dynamics}Average versus kink velocity}
\begin{figure}
	\includegraphics[width=1.0\linewidth]{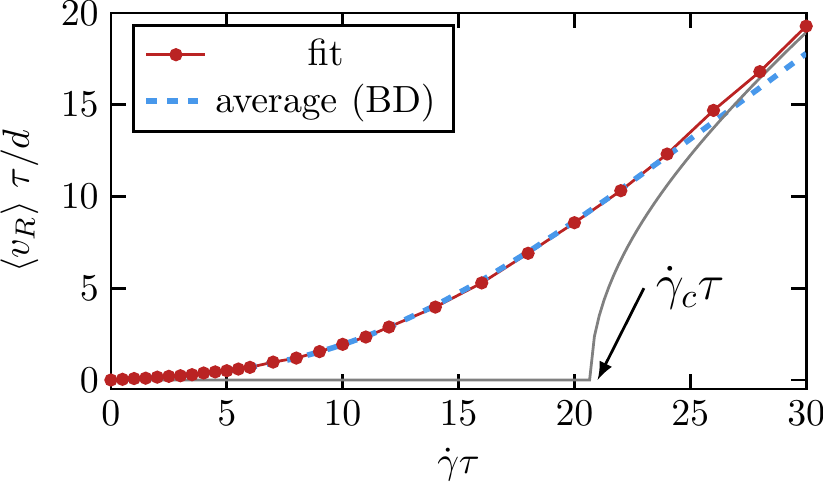}
	\caption{(Color online) The relative velocity of the top layer (blue dashed) and the approximation via the ansatz \eref{estimated relative velocity top layer} (red). The velocity of the ideal (crystalline) top layer \eref{equation of motion simple model} and its depinning transition are included as a reference.}
	\label{fig:estimated relative velocity binary}
\end{figure}
The results discussed in \sref{single particle and cluster dynamics} suggest that kinks represent the only mechanism leading to the mean particle transport of the top layer.
Motivated by the corresponding formula in the FK model [see \eref{FK velocity relation}], we thus propose to describe the average velocity in our system as
\begin{equation}\label{eq:estimated relative velocity top layer}
	\av{v_{R}} = \alpha\lr{\shearrate} \frac{N_{K}}{N_L} \av{v_{kink}}\text{,}
\end{equation}
where $\alpha$ is a (shear-rate dependent) factor of proportionality and $N_L$ is the number of particles in the top layer.
Of course, this relation is expected to hold only for shear rates, where the shear forces are not yet sufficient to introduce free sliding of the top layer and kinks are indeed the main transport mechanism.
In order to estimate this range of validity of \eref{estimated relative velocity top layer}, we calculated the critical shear rate $\shearrate_c \tau \approx 21$ for the depinning transition of the top layer (assuming that the latter is perfectly crystalline for $\shearrate<\shearrate_c$) by using the model presented in \sref{mapping to shear-driven system}.

Numerical results of this analysis are shown in \fref{estimated relative velocity binary}.
Fitting $\av{v_R}$ according to \eref{estimated relative velocity top layer} we find that  $\alpha\lr{\dot{\gamma}} \approx 1.17 + 0.08 \dot{\gamma}\tau$.
The agreement between \eref{estimated relative velocity top layer} and the true kink velocity is particularly good in the range $\shearrate < \shearrate_c$.
Only for "supercritical" shear rates (i.e., $\shearrate > \shearrate_c$), we observe significant deviations.
Here, the top layer is completely melted and the system displays collective motion of the particles in large density waves.
This is obviously strongly different from the transport mechanism of kinks.
Finally, in comparison to the FK model, we find that the factor of proportionality ($\alpha\lr{\shearrate}$) is weakly shear-rate dependent, corresponding to a small thermal drift of the top layer.
However, especially for small shear rates, this drift can be neglected, reflecting that kinks are indeed the dominant transport mechanism.
Very similar results are obtained for somewhat larger sedimentation strengths \footnote{
  For example, for $\sedimentationStrength\, d^4/k_B T = 350$, we find that $\alpha\lr{\shearrate} \approx 1.17 + 0.07\shearrate\tau$.
}.
\section{\label{sec:conclusion}Conclusion}
Using BD simulations and an analytical approach we have studied the dynamical behavior of three types of colloidal films under planar shear flow.
Focusing on high densities and strong confinement, where the colloids arrange in two or three layers with (squarelike) crystalline order, the shear-induced dynamical behavior is similar to that of colloidal monolayers driven over a periodic substrate potential \cite{Hasnain2013, Bohlein2012}.
In particular, the symmetric (one-component) bilayer system displays a depinning transition, where the layers are "pinned" to each other up to a critical shear rate \cite{Vezirov2013}.
A similar depinning transition is also observed for the symmetric (one-component) trilayer system.
Interestingly, this does not hold for the asymmetric (two-component) trilayer system, which is characterized by a mismatch of the effective lattice constants in the top and the two bottom layers.
In this system, the top layer is never fully pinned, rather we observe the formation of kinklike defects reminiscent of the FK model \cite{Braun1998}.

From a conceptual point of view, one key result of our study is that the dynamics of the symmetric systems can be mapped to the motion of a single particle driven over an effective periodic substrate potential.
The resulting effective model can be solved analytically and yields a prediction of the critical shear rate for the depinning transition.
For the bilayer system, both the resulting average velocity of the layers and the shear stress are in good qualitative agreement with the BD simulation results.
Further, the mapping procedure reveals the relation between the critical shear rate and important system parameters such as the strength of the pair interactions and the width of the confinement.
For the symmetric trilayer system, the critical shear rate is overestimated in the sense that the effective model cannot describe the laned state which occurs in the real system between the crystalline and the melted state.
Still, the model predicts nearly correctly the onset of melting.

Another main result of our study is the observation of local transport via kinklike density excitations in the asymmetric trilayer system.
For small shear rates, the kinks provide the main mechanism for particle transport in the top layer.
The average velocity of the layer is then proportional to their average velocity times the number of kinks.
The factor of proportionality is weakly shear rate dependent, which we interpret as a small thermal drift due to the noise.
Interestingly, the antikink-like defects do not contribute to the particle transport, rather they are stationary relative to the top layer.
This is in contrast to the FK model and can be explained by deformations of the bottom layers in response to the locally reduced density in the top layer.

Similar to previous studies \cite{Vezirov2013, Vezirov2015}, we here employed a set of system parameters pertaining to a realistic system of charged silica particles \cite{Klapp2007, Klapp2008, Zeng2011}.
Thus, our predictions can, in principle, be tested by experiments.
In this context we note that the presence of a solvent can induce hydrodynamic interactions between the colloidal particles, which are neglected in our model.
Considering experimental studies confirming the solidlike response of strongly confined fluids \cite{Gee1990} and local transport via kinklike defect structures \cite{Bohlein2012}, we expect these interactions to affect the timescales, but not the overall behavior of the system.

In addition to a direct comparison to experiments, it would be very interesting to investigate the shear-induced dynamics of confined films for wall distances corresponding to a hexagonal or disordered equilibrium configuration as well as the dynamics of thicker binary crystalline films \cite{Horn2014,Assoud2008}.
Further interesting aspects are the impact of oscillatory shear flow and of structured walls (which can influence the crystalline structure \cite{Besseling2012, Wilms2012}) on the dynamics of the system.
We also note that there is increasing interest concerning the interplay of shear flow and strong confinement in glasslike colloidal systems (see \cite{Chaudhuri2013} for a corresponding molecular dynamics study with a much wider slit-pore width).

Especially for the latter systems, the local particle transport in defect structures might be key.
To this end, it seems vital to better understand the relation between the structural properties of kinks (as well as of antikinks) and their dynamics.
A first step in this direction would be to investigate the dependence of the defect velocity on the size of the defects, as well as corresponding relaxational time scales.
Work in these directions is in progress.
\section{Acknowledgments}
This work was supported by the Deutsche Forschungsgemeinschaft through SFB 910 (Project No. B2).
\appendix
\section{\label{sec:appendix:optimal path}Optimal path}
\begin{figure}
	\includegraphics[width=1.0\linewidth]{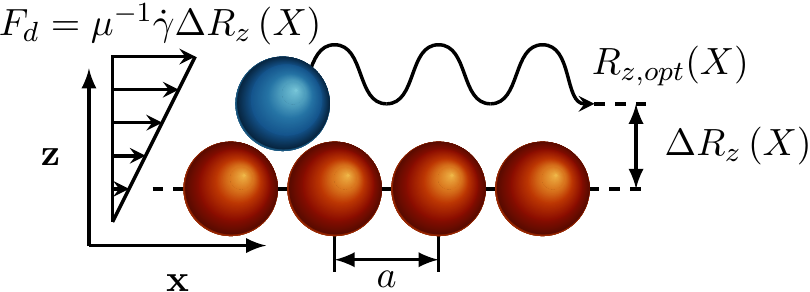}
	\caption{(Color online) Sketch of the optimal path of the center of mass of the top layer (blue particle) during its motion over the bottom layer (red particles).}
	\label{fig:optimalPosition}

\end{figure}
To describe the motion of the top layer in the driven system we define an "optimal" path $R_{z,opt}\lr{X}$.
The latter describes the motion of the center of mass of the top layer assuming that the bottom layer is in its equilibrium configuration (see \fref{optimalPosition}). Specifically, we define $R_{z,opt}\lr{X}$ via the condition
\begin{equation}\label{eq:optimal path}
	F_{z}^{inter,bot}\lr{R_{z,opt}} + F_{z}^{wall}\lr{R_{z,opt}} = 0 \text{.}
\end{equation}
According to \eref{optimal path}, the $z$-position of the top layer is adjusted such that the force from the bottom layer and the confinement in $z$-direction is balanced for all displacements in $x$-direction.
This ansatz is reasonable when we assume that the relaxational time scale of the top layer in $z$-direction, $\tau_z \ll a / \av{v_R}$, where $a$ is the lattice constant of the bottom layer, is very small as compared to the typical time scale of the sliding dynamics.

In order to get simple analytic expressions for $F_d$ and $F_{sub}$ [see  \eref{driving force} and (\ref{eq:reduced substrate force})], we calculate numerically the corresponding Fourier series and neglect all higher harmonics. This yields
\begin{align}
	\label{eq:effective substrate force}
	F_{sub}\lr{X} &\approx F_{max}\sin \lr{\frac{2\pi}{a} X  }\\
	\label{eq:layer distance optimal path}
	F_d\lr{X} &= \frac{\shearrate}{\mu} \Delta R_z\lr{X} \nonumber \\
						&\approx \frac{\shearrate}{\mu} \LR{ Z_A \cos\lr{\frac{2\pi}{a} X } + Z_0 } \text{,}
\end{align}
where $F_{max}$ is the amplitude of the substrate force, $Z_A$ is the amplitude of $\Delta R_z\lr{X}$ and $Z_0 = a^{-1} \int_{0}^{a} \Delta R_z\lr{X} dX$ is the mean layer distance.
Equation (\ref{eq:effective substrate force}) and (\ref{eq:layer distance optimal path}) imply that the period of the spatial oscillations is the same in both quantities.
\section{\label{sec:appendix:trajectory}Trajectories within the effective model}
\begin{figure}
	\includegraphics[width=1.0\linewidth]{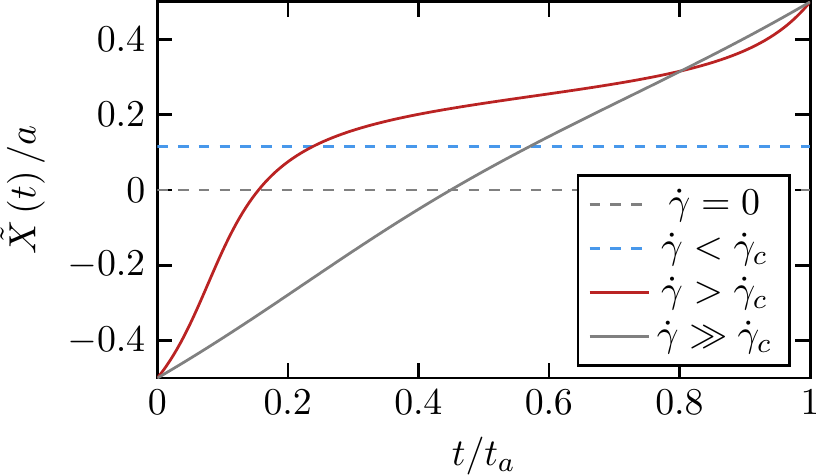}
	\caption{(Color online) Illustration of the long-time trajectories given in \eref{long-time position} for different shear rates in one cycle (with period $ t_a = a / \av{v_R}$).}
	\label{fig:steady state position bilayer}
\end{figure}
The equation of motion (\ref{eq:equation of motion simple model}) presented in \sref{driven monolayers} can be simplified by using trigonometric identities, yielding
\begin{equation}\label{eq:equation of motion simple model shifted}
	 \dot{X} = \tilde{F}_{max} \sin\lr{\frac{2\pi}{a} X+\Phi} + \shearrate Z_0.
\end{equation}
Here, the $X$-dependence of the driving force is accounted for by the rescaled substrate force $\tilde{F}_{max}$ and the constant phase shift $\Phi$, which are given by
\begin{align}\label{eq:rescaled substrate force and shift}
	\tilde{F}_{max} &= \sqrt{\mu^2 F_{max}^2 + \shearrate^2 Z_A^2}\text{,} \\
	\Phi &= \tan^{-1}\lr{ \frac{\shearrate Z_A}{\mu F_{max}} }.
\end{align}
Substituting $\bar{X} = 2\pi X / a +\Phi$ we arrive at the standard Adler equation \cite{Adler1946} for $\bar{X}$, which can be solved analytically.
The solution for $\bar{X}$ reads
\begin{equation}\label{eq:trajectories optimal path}
	 \bar{X}\lr{t} = 2\tan^{-1}\lr{ \frac{ \av{v_R} \tan\lr{\frac{ \pi\av{v_R}}{a}\lr{t+t_0} } -  \tilde{F}_{max} }{\shearrate Z_0} },
\end{equation}
where $\av{v_R}$ is the average velocity given in \eref{relative velocity for optimal path}, and $t_0$ is a constant of integration.

We now focus on the long-time solutions (thus, neglecting relaxational dynamics) defined by $\tilde{X}\lr{t} = \lim_{t\to\infty}\bar{X}\lr{t}$.
At long times, one has
\begin{equation}\label{eq:long time limit}
 \lim_{t\to\infty} \lr{t+t_0} = t \text{,}
\end{equation}
that is, the initial time $t_0 \ll t$ can be neglected.
We consider the two cases $\shearrate \leq \shearrate_c$ and $\shearrate>\shearrate_c$ separately.
Using \eref{long time limit} for the case $\shearrate \leq \shearrate_c$ and substituting $\av{v_R} = i \av{v_R^*}$, with the complex conjugated average velocity $\av{v_R^*}\in\mathbb{R} \quad\forall\; \shearrate \leq \shearrate_c $, yields
\begin{align}\label{eq:stationary limit}
	\lim_{t\to\infty} \tan\lr{ i \frac{\pi \av{v_R^*}}{a} t } &= i\, \lim_{t\to\infty}  \text{tanh}\lr{ \frac{\pi \av{v_R^*}}{a} t }\nonumber\\
	 &= i \text{,}
\end{align}
where $i$ is the imaginary unit and we used the identity $\text{tanh}\lr{x} = -i\, \tan \lr{i x}$.
Inserting Eqs. (\ref{eq:long time limit}) and (\ref{eq:stationary limit}) into \eref{trajectories optimal path} and doing the same analysis (yet with the real velocity) for $\shearrate>\shearrate_c$, the long-time solutions read
\begin{widetext}
\begin{equation}\label{eq:long-time position}
	\tilde{X}\lr{t}=
	\begin{cases}
		2\tan^{-1}\LR{  -\lr{\av{v_R^*} + \tilde{F}_{max} } / \shearrate Z_0 }&,  \shearrate \leq \shearrate_c\\[10pt]
		2\tan^{-1}\LR{ \lr{ \av{v_R} \tan\lr{\frac{ \pi\av{v_R}}{a} t } -  \tilde{F}_{max} } / \shearrate Z_0 }&,  \shearrate > \shearrate_c

	\end{cases}.
\end{equation}
\end{widetext}
The two solutions (for representative parameters) are plotted in \fref{steady state position bilayer}.
For $\shearrate \leq \shearrate_c$, the layer is locked, yielding a constant displacement $\tilde{X}$.
The nonzero value of $\tilde{X}$ for $\shearrate\neq 0 $, $\shearrate \leq \shearrate_c$ reflects the elastic displacement due to the shear.
Increasing the shear rate to supercritical values, $\shearrate > \shearrate_c$, we find an oscillatory running state, which is characterized by fast motion from one lattice position to the next and a slow "build-up phase" in between.
This motion transitions into an uniform free sliding (i.e., quasi-linear increase of $\tilde{X}$) for very large shear rates, $\shearrate \gg \shearrate_c$.
%
%
% \bibliography{/home/sgerloff/Documents/Bibliography/Bibliography.bib}
\bibliography{citation.bib}
% \bibliography{citation2.bib}
% \input{main.bbl}

\end{document}